\documentclass[letterpaper, 10 pt, conference]{ieeeconf}  

\IEEEoverridecommandlockouts                              

\usepackage{epsfig} 
\usepackage{amsmath} 
\usepackage{amssymb}  

\usepackage{multirow}
\usepackage{algorithm}
\usepackage{algpseudocode}

\title{\LARGE \bf
Generalizable Deep Learning Method for Suppressing Unseen and Multiple MRI Artifacts Using Meta-learning
}

\author{Arun Palla$^{1,\star}$, Sriprabha Ramanarayanan$^{1,2,\star}$, Keerthi Ram$^{2}$, and Mohanasankar Sivaprakasam$^{1,2}$
\thanks{$^{1}$Indian Institute of Technology Madras (IITM), 600036, India}%
\thanks{$^{2}$Healthcare Technology \& Innovation Centre (HTIC), 600113, India}%
\thanks{$^{\star}$Equal contribution. Author is available at arunpalla97@gmail.com}
}

\begin{document}

\maketitle
\thispagestyle{empty}
\pagestyle{empty}

\begin{abstract}

Magnetic Resonance (MR) images suffer from various types of artifacts due to motion, spatial resolution, and under-sampling. Conventional deep learning methods deal with removing a specific type of artifact, leading to separately trained models for each artifact type that lack the shared knowledge generalizable across artifacts. Moreover, training a model for each type and amount of artifact is a tedious process that consumes more training time and storage of models. On the other hand, the shared knowledge learned by jointly training the model on multiple artifacts might be inadequate to generalize under deviations in the types and amounts of artifacts. Model-agnostic meta-learning (MAML), a nested bi-level optimization framework is a promising technique to learn common knowledge across artifacts in the outer level of optimization, and artifact-specific restoration in the inner level. We propose curriculum-MAML (CMAML), a learning process that integrates MAML with curriculum learning to impart the knowledge of variable artifact complexity to adaptively learn restoration of multiple artifacts during training. Comparative studies against Stochastic Gradient Descent and MAML, using two cardiac datasets reveal that CMAML exhibits (i) better generalization with improved PSNR for 83\% of unseen types and amounts of artifacts and improved SSIM in all cases, and (ii) better artifact suppression in 4 out of 5 cases of composite artifacts (scans with multiple artifacts).
\newline

\indent \textit{Clinical relevance}— Our results show that CMAML has the potential to minimize the number of artifact-specific models; which is essential to deploy deep learning models for clinical use. Furthermore, we have also taken another practical scenario of an image affected by multiple artifacts and show that our method performs better in 80\% of cases.
\end{abstract}

\section{INTRODUCTION}

Image restoration of artifact affected scans is a challenging problem in Magnetic Resonance Imaging (MRI) that is crucial to allow the radiologist to arrive at a better diagnosis. Common MRI artifacts arise from movement of anatomical structures, and trade-off among spatial and temporal resolution, scan time, and signal-to-noise ratio (SNR) \cite{ismail2022cardiac}. Conventional methods for artifact suppression like compressed sensing \cite{cs_mri} treat artifacts as aliasing or blur and use non-linear optimization solvers to iteratively recover the image. Recently, deep learning methods have shown faster inference and promising results in image restoration, and can be effectively adopted for enhancing degraded scan quality.

Existing deep-learning (DL) techniques address the removal of a single type of artifact. CINENet \cite{Kstner2020CINENetDL} performs artifact restoration of only under-sampled Cardaic MR (CMR). Several deep neural network architectures \cite{lyu2021cine}, \cite{xia2021super}, \cite{masutani2020deep} are developed specifically for motion correction or super-resolution. However, as MRI is affected by diverse and multiple sources of artifacts, training a model separately for each type of artifact incurs large computational requirements. Secondly, these models cannot be scaled to various unseen acquisition settings, which makes it hard to deploy on commercial MRI machines. Moreover, a single model that learns a common knowledge across various MRI artifacts \cite{univusmri}, \cite{ramanarayanan2020mac} helps to generalize to unseen artifacts with deviated amounts of degradation levels at test time (e.g., under-sampling with unseen acceleration factors different from training data). We consider the problem of restoring artifact affected MRI images for various types of artifacts in a single DL model with a training process driven by learning a set of latent representation common across artifacts.
\begin{figure}[t]
    \centering
    \includegraphics[width=\linewidth]{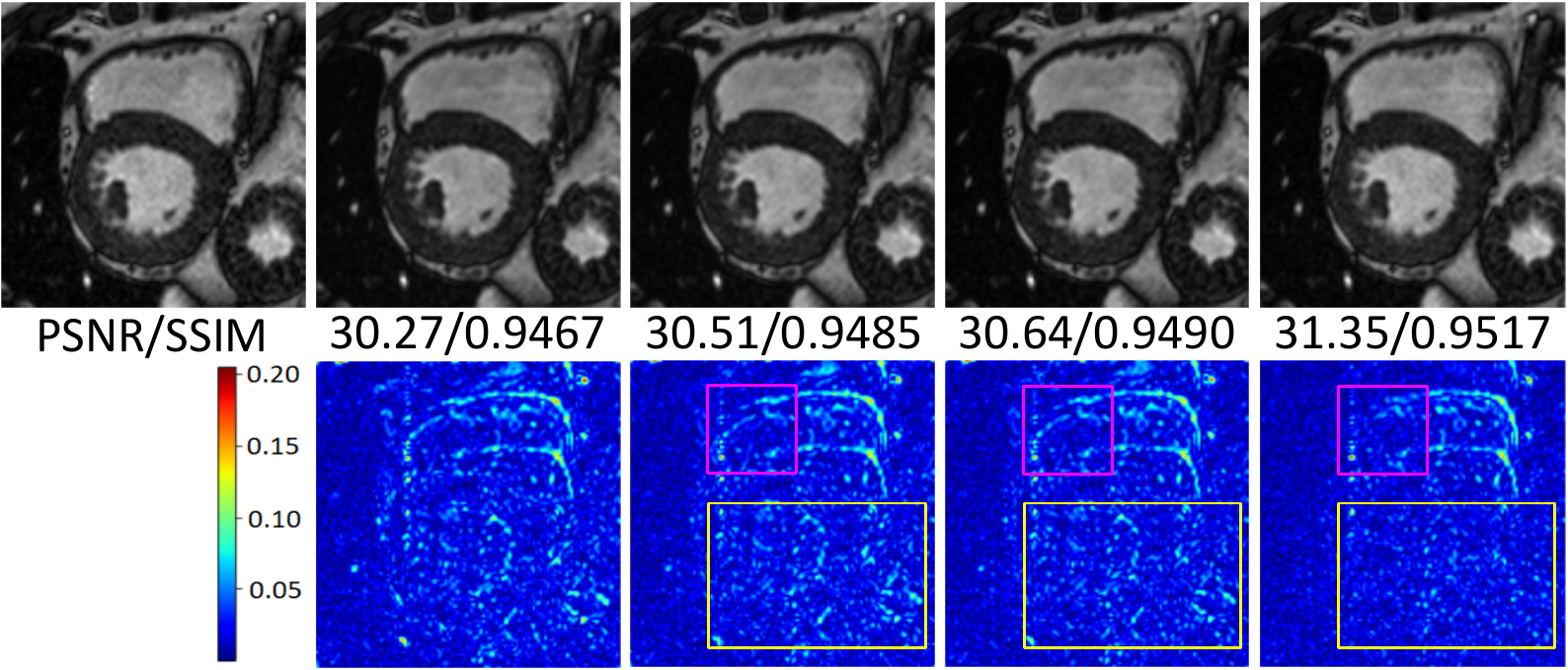}
    \caption{Qualitative results on motion affected cardiac scan. Top row (left to right): Target image, motion artifact image, predictions of joint training, MAML and CMAML. Bottom row: Reconstruction residue images with respect to target.
    }
    \label{fig:Motion artifact visualization}
\end{figure}

\begin{figure*}
    \centering
    \includegraphics[width=\linewidth]{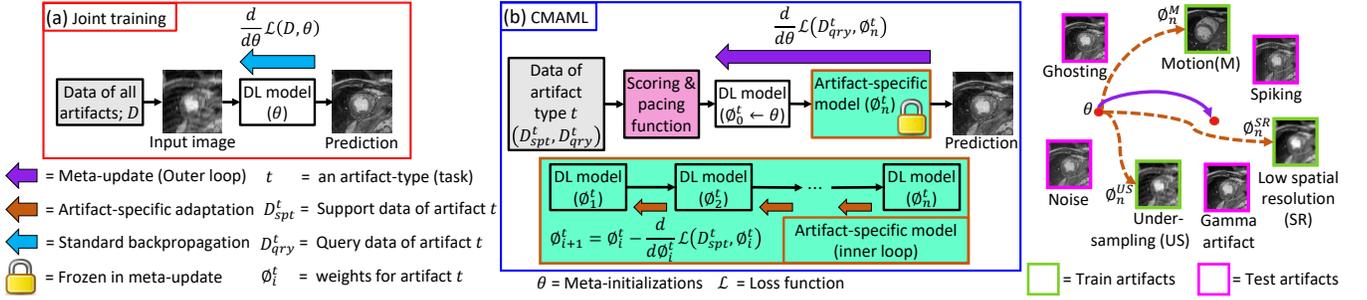}
    \caption{Graphical abstract of (a) Joint training and (b) CMAML for a given neural network. Joint training strategy uses a single-level standard backpropagation (blue arrow) and combines all train samples. CMAML training process adopts a bi-level optimization on tasks defined by the type of artifact. The two levels of optimization are: 1) meta-update (outer loop shown as dark violet arrow) and 2) artifact-specific adaptation via few gradient descent steps (inner loop with brown arrow). (Right) Visualization of the two levels of optimization in CMAML in the task space of various artifacts.
    }
    \label{fig:Graphical abstract and task space visualization}
\end{figure*}

Traditional training methods, hereafter called as \emph{Joint training} (or) \emph{ Stochastic Gradient Descent (SGD)}, combines all artifact data in a single level of optimization as shown in Fig. \ref{fig:Graphical abstract and task space visualization}a. Such methods might not include additional nuances of artifacts (like artifact-type and artifact-amount) that are essential to drive the discriminative artifact-specific restoration during training. In our work, we propose a training process that exhibits two key attributes: 1) exploits the artifact-invariant information that is characterized by the common knowledge shared across images of different artifacts and 2) incorporates discriminative artifact-specific fine-tuning at train time; to achieve restoration of various degradations, all in a single model.

Inspired by the recently emerged nested bi-level (inner level and outer level) optimization framework of Model-agnostic meta-learning (MAML) \cite{finn2017model} that is based on a set of observed tasks, we propose a learning model for suppressing multiple MRI artifacts, that facilitates common knowledge in the outer level and performs artifact-specific restoration in the inner level as shown in Fig. \ref{fig:Graphical abstract and task space visualization}b. In particular, the artifact-specific model in the forward-pass of our training process in Fig. \ref{fig:Graphical abstract and task space visualization}b (green block) effectively promotes a generalizable model $(\theta)$, based on artifact type, even in the unseen test degradations. In our MAML framework, we pose each type of artifact as a task. We extend the MAML framework further by incorporating curriculum learning \cite{hacohen2019power} (CL), in order to provide additional knowledge of artifact-complexity during training. We introduce the level of complexity in CL based on the ill-posedness of various artifacts. For instance, in under-sampling artifact, the level of complexity can be directly related to the ill-posedness characterized by the amount of acceleration during MRI acquisition. Thus, for two scans where one scan is $3\emph{x}$ accelerated and the other is $5\emph{x}$, the former task is less complex than the latter. The proposed learning model can embed representations based on the nuances in artifacts for a given neural network architecture.
Our contributions are:

1. We propose curriculum-based model-agnostic meta-learning called CMAML, a training method that performs image restoration of multiple types of artifact affected MRI scans in a single model. The proposed method establishes an artifact-type as a task and associates the amount of MRI artifact with the task-complexity.

2. Under unseen artifacts that are different from train artifacts, we have compared our model with other learning baselines, namely joint training and MAML. Results show that our model exhibits better performance over other methods in all of the unseen artifacts.

3. We evaluate the methods on composite artifact data (same scan with multiple artifacts). Comparative results illustrate that our method outperforms in 80\% of artifacts.

\section{METHODOLOGY}
\label{sec:methods}

The data acquisition forward model of the artifact restoration problem \cite{yang2017dagan} can be formulated as:
\begin{align}
 \label{Forward model}
 \mathcal{A}x+\epsilon = y
\end{align}

where, $x$ $\in$ $\mathbb{C}^\emph{N}$ denotes the desired image, $y$ $\in$ $\mathbb{C}^\emph{M}$ is the measurement from the MRI scanner, $\epsilon \in \mathbb{C}^\emph{M}$ is the noise and $\mathcal{A} : \mathbb{C}^\emph{N} \rightarrow \mathbb{C}^\emph{M}$ represents the forward operator of MRI acquisition that causes artifacts. Artifact restoration is ill-posed as the problem is under-determined ($M << N$) and the operator $\mathcal{A}$ is ill-conditioned. The restoration of MRI artifact scans is achieved by introducing an a-priori knowledge of $x$ into the unconstrained optimization \cite{yang2017dagan}: 
\begin{align}
 \label{Optimization formulation}
 \underset{x} {\operatorname{min}} \quad ||\mathcal{A}x - y||_{2}^2 + \mathcal{R}(x)
\end{align}

where, $||\mathcal{A}x - y||_{2}^2$ is the data fidelity term and $\mathcal{R}$ is a regularization term.

Deep learning-based MRI image restoration involves training a DL model using a single-level optimization on the average loss of all observed artifact data. This supervised joint training can be formulated as:
\begin{align}
 \label{Joint-training formulation}
 \theta^{*} = \underset{\theta} {\operatorname{argmin}}  \displaystyle \mathop{\mathbb{E}}_{(x_{in},x)\in \bigcup\limits_{i}^{} \mathcal{D}_{i}}[||x - f(x_{in};\theta) ||_{2}^2]
\end{align}

where, $\mathcal{D}_{i}$ represents the dataset of artifact \emph{i}, consisting of ground truth image \emph{x} and its corresponding degraded image $x_{in}$. Here, \emph{f} is a DL model parameterized by $\theta$. Unlike iterative methods in Eq. \ref{Optimization formulation}, statistical expectation ($\displaystyle \mathop{\mathbb{E}}$) analysis in Eq. \ref{Joint-training formulation} infers an optimal parameter set $\theta^{*}$. Artifacts considered for training are motion, super-resolution (SR) and under-sampling (US), each defined with a corresponding operator as:

(i) \(\mathcal{A}_{M}(x) = \mathcal{F}^{-1}[\sum_{j=l-s}^{l+s} \psi_{j}(\mathcal{F}(x_{j}))]\), where $\mathcal{F}$ and $\mathcal{F}^{-1}$ denote 2D Fourier and inverse transforms respectively, $\psi$ selects k-space data lines of $2s+1$ frames of CMR \cite{lyu2021cine}.

(ii) \(\mathcal{A}_{SR}(x) = \mathcal{U}(\mathcal{D}(x))\), where $\mathcal{D}$ and $\mathcal{U}$ denote downsampling and upsampling operations respectively.

(iii) \( \mathcal{A}_{US}(x) = \mathcal{F}^{-1}(\mathcal{M}\circ \mathcal{F}(x))\) where $\circ$ indicate Hadamard product and $\mathcal{M}$ is an under-sampling mask.

MAML involves partitioning each task's data into support ($x^{t}_{in,spt},x_{spt}$) and query ($x^{t}_{in,qry},x_{qry}$) samples. The parameters of DL model, $\theta$, in Fig. \ref{fig:Graphical abstract and task space visualization}b are called \emph{meta-initializations}. For every task \emph{t}, we use support samples to perform a few gradient-descent steps (\emph{adaptation}) from meta-initializations to obtain task-specific parameters ($\phi_{t}$). Loss of task-specific parameters on query data is aggregated over all train tasks ($\mathcal{T}$), to provide supervision for meta-initializations. The training method of MAML \cite{hospedales2021meta} is:
\begin{align}
\label{MAML training formulation}
 \theta^{*} = \underset{\theta} {\operatorname{argmin}} \sum_{t\in \mathcal{T}} ||x_{qry} - f(x^{t}_{in,qry};\phi_{t}) ||_{2}^2 \\ s.t \quad \phi_{t} = \underset{\theta} {\operatorname{argmin}} ||x_{spt} - f(x^{t}_{in,spt};\theta) ||_{2}^2 \quad \forall t\in \mathcal{T}
\end{align}

\begin{algorithm}[b!]
\caption{CMAML algorithm}
\label{CMAML algo}
\begin{algorithmic}[1]
\Require $\alpha$, $\beta$ and \( \mathcal{T} \): Train tasks
\Require \( S(\mathcal{T}) \): Task scoring-function
\Require \( P(\mathcal{T},.) \): Task pacing-function
\State Randomly initialize $\theta$
\State Sorted tasks: $\mathcal{T}_{sorted}$, in ascending order by \( S(\mathcal{T}) \)
\For{$e = 1\ to\ max\_epochs$}
\State Infer a task mini-batch: $\mathcal{T}_{batch}$, U = \( P(\mathcal{T}_{sorted},e) \)
\State $\mathcal{L}_{meta} = 0$
\For{each training task $t$ in \( \mathcal{T}_{batch} \)}
\State Sample a mini-batch of support data

\hskip0.9em$D_{spt}^{t}=\{ X_{in,n}^{t},X_{n} \}_{n=1}^{N_{spt}}$
\State Initialize $\phi_{t} \leftarrow \theta$
\For{$u = 1 \ to \ U$} \Comment{Artifact-specific adaptation}
\State $\mathcal{L}_{1} = L(\phi_{t}, D_{spt}^{t})$
\State $\phi_{t} \leftarrow \phi_{t}-\alpha \nabla_{\phi_{t}}\mathcal{L}_{1}$ 
\EndFor
\State Sample a mini-batch of query data

\hskip0.9em$D_{qry}^{t}=\{ X_{in,n}^{t},X_{n} \}_{n=1}^{N_{qry}}$
\State $\mathcal{L}_{meta} \leftarrow \mathcal{L}_{meta} + L(\phi_{t}, D_{qry}^{t})$
\EndFor
\State $\theta \leftarrow \theta - Adam[\beta,\mathcal{L}_{meta}]$ \Comment{Meta-update}
\EndFor
\end{algorithmic}
\end{algorithm}

Here, in our work of MAML-based artifact image restoration, we consider an artifact-type as a task. On every artifact's support samples, we perform adaptation to result in an artifact-specific model that is characterized by weights $\phi_{t}$ and it constitutes one-level of optimization (inner level, \emph{steps 6 to 12} in Algorithm \ref{CMAML algo}) in the bi-level MAML framework. The adaptation on an artifact is shown in the green block of Fig. \ref{fig:Graphical abstract and task space visualization}b. After adapting on support samples of all train artifact data, we aggregate the loss incurred by every artifact-specific model using query samples. In the second-level of optimization (outer level, \emph{steps 13 to 16} of Algorithm \ref{CMAML algo}), we backpropagate the total loss to update the meta-initializations and thus completing one end-to-end iteration in the course of training. During the outer level update, the parameters of the artifact-specific model are frozen to perform optimization on only the meta-initializations as shown in Fig. \ref{fig:Graphical abstract and task space visualization}b. However, additional nuances in the complexity of various configurations within a type of artifact can embed representations that generalize better for unseen artifacts in test data.

In Curriculum-learning (CL) \cite{hacohen2019power}, there is 1) a scoring-function that estimates the complexity of a data-sample based on the difficulty of predicting the target associated with the input and 2) a pacing-function that schedules the inclusion of data-samples to DL model across epochs as the training progresses.
During the initial epochs of training in CL, pacing-function provides only low-complexity data-samples based on scoring function to the DL model and as the training progresses, medium to high complexity data-samples are cumulatively included.

\begin{figure}
    \centering
    \includegraphics[width=0.8\linewidth]{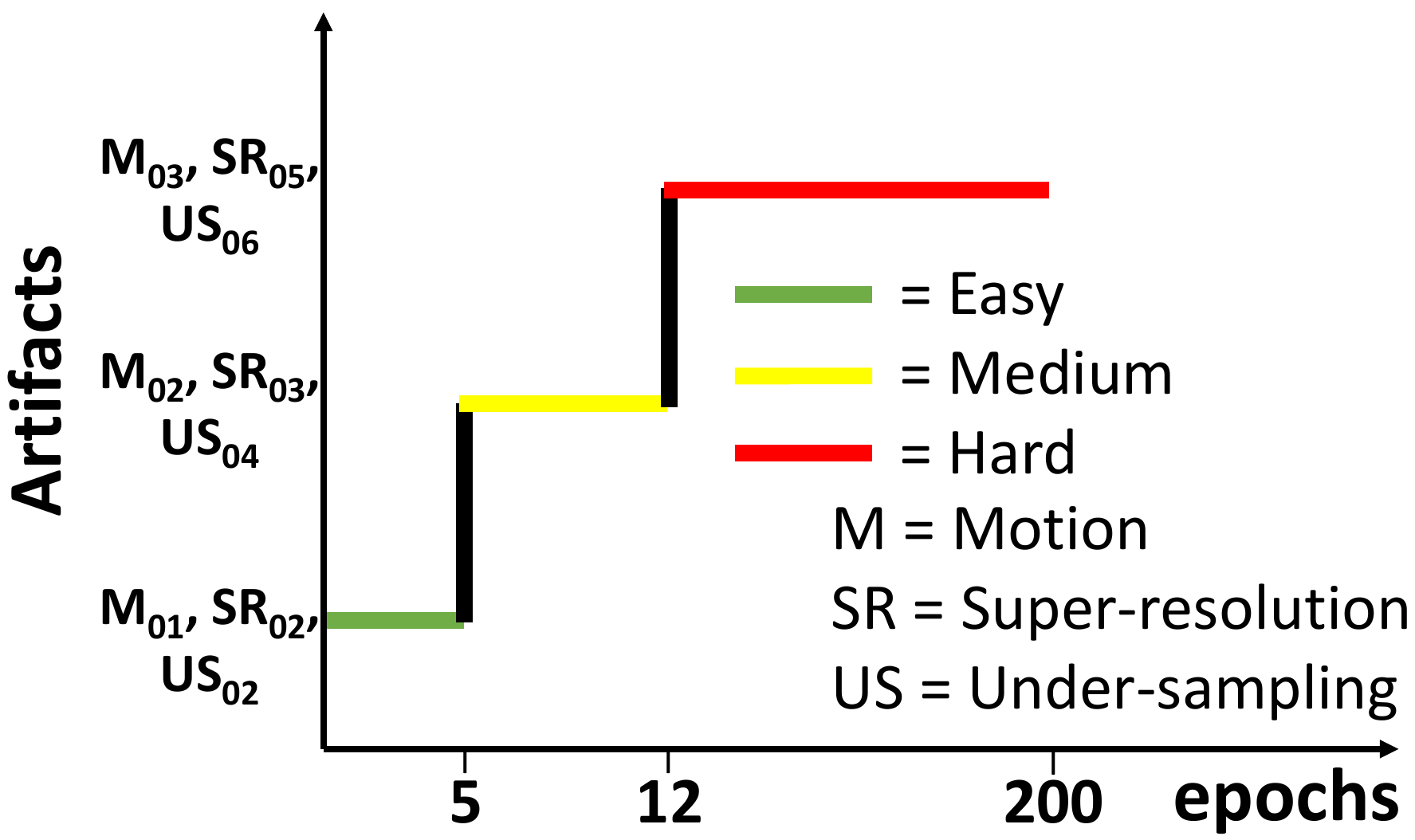}
    \caption{Pacing function. $US_{02}$ indicates 2x under-sampling. Tasks are cumulatively added during training.}
    \label{fig:Pacing function}
\end{figure}

Here, we further incorporate CL into MAML framework. Unlike CL, we define complexity over a task (Fig. \ref{fig:Graphical abstract and task space visualization}b, pink block) and not on data-samples. The scoring-function of a task is parameterized by the ill-posedness of the corresponding forward operator $\mathcal{A}$. The pacing-function we used is a simple step-function of epoch as in CL \cite{hacohen2019power}, as shown in Fig. \ref{fig:Pacing function}. The training process of CMAML is described in Algorithm \ref{CMAML algo} and various notations are explained in Section \ref{sec:implementatin details}.

\begin{figure*}[ht]
    \centering
    \includegraphics[width=0.8\linewidth]{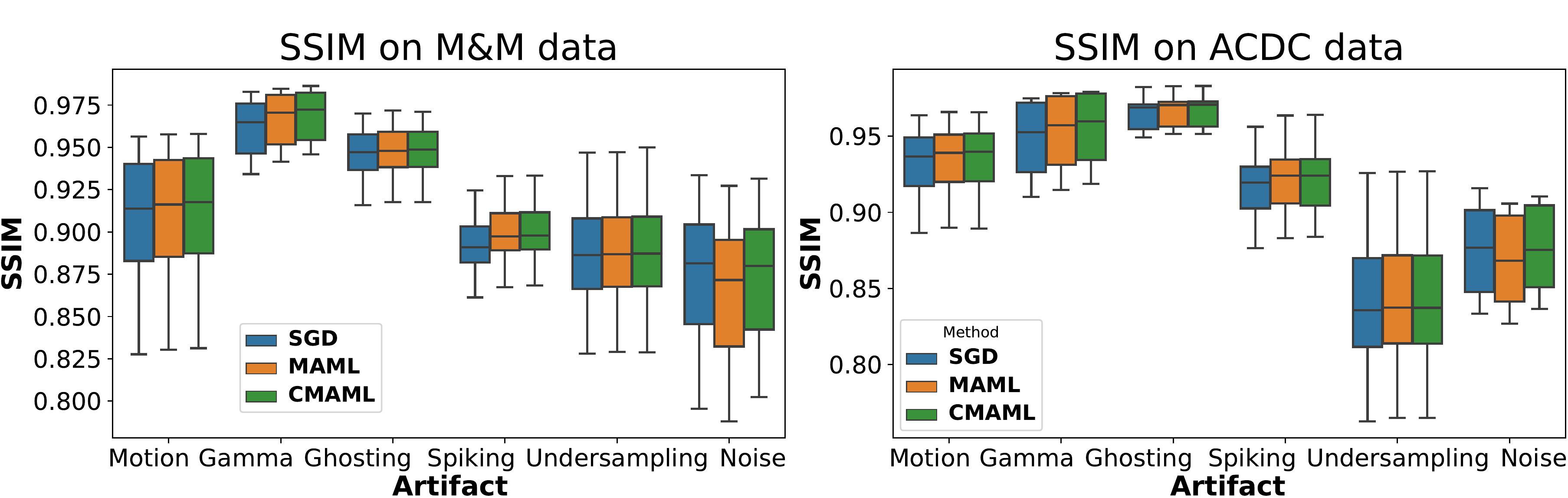}
    \caption{
    Comparative analysis with respect to SSIM metric on unseen artifact data.
    }
    \label{SSIM boxplot}
\end{figure*}
\begin{figure}[t]
    \centering
    \includegraphics[width=\linewidth]{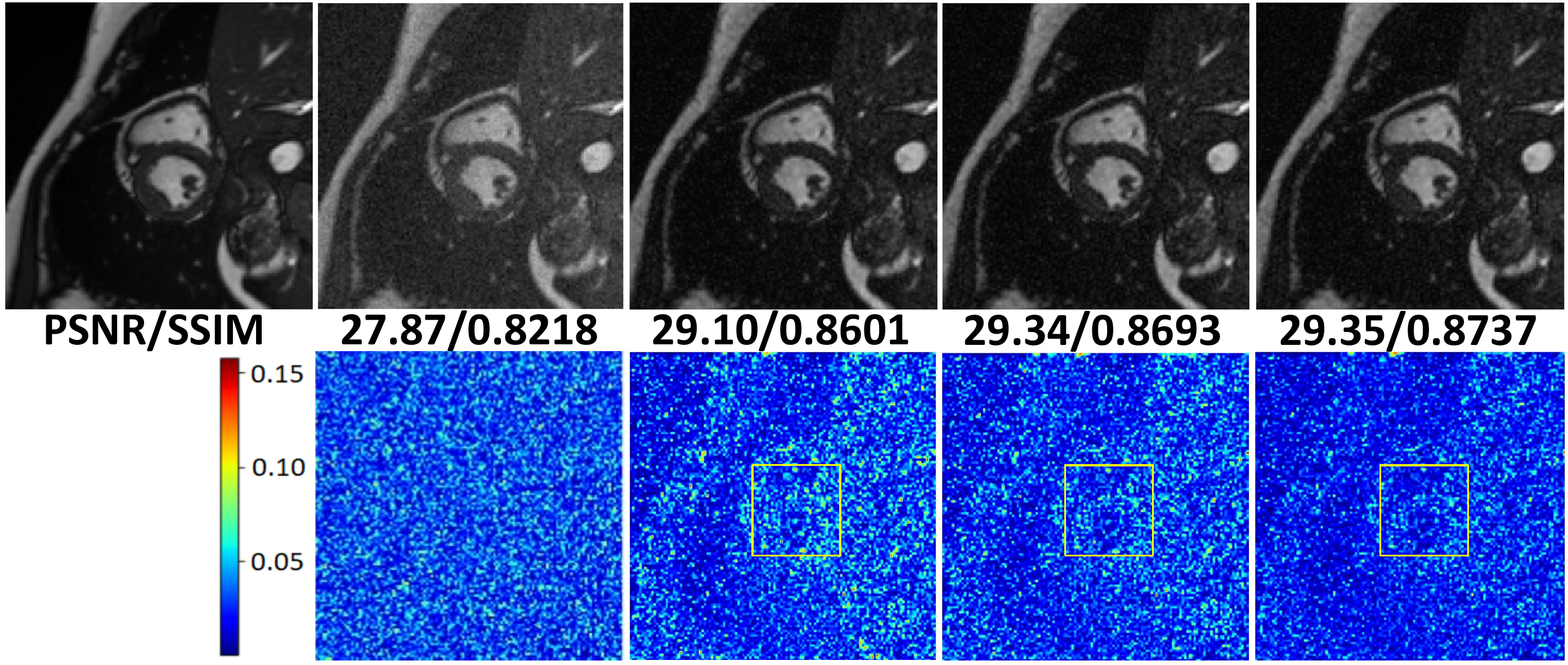}
    \caption{Qualitative results on unseen artifact data. Top row (left to right): Target image, noise artifact image, predictions of joint training, MAML and CMAML. Bottom row: Reconstruction residue images with respect to target
    }
    \label{fig:Noise artifact visualization}
\end{figure}
\begin{figure}
    \centering
    \includegraphics[width=\linewidth]{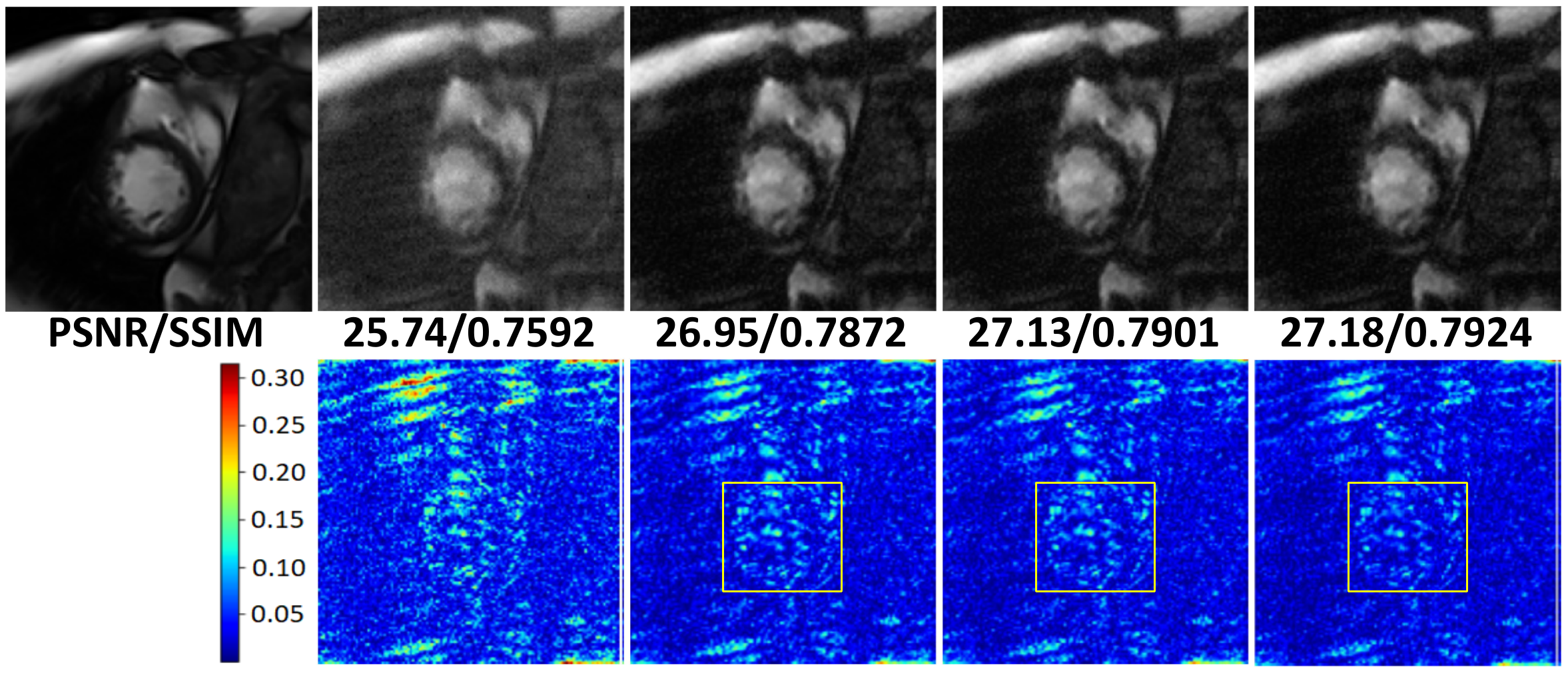}
    \caption{Qualitative results on composite artifact data. Top row (left to right): Target image, image corrupted by undersampling and noise artifact, predictions of joint training, MAML and CMAML. Bottom row: Reconstruction residue images with respect to target.
    }
    \label{fig:Composite artifact visualization}
\end{figure}
\begin{table}[t]
\scriptsize
\centering
\caption{Quantitative results of various unseen artifacts. US denotes Under-sampling. Bold figures indicate the best scores.}
\label{Tabulation results}
\begin{tabular}{|c|c|c|c|c|}
\hline
\multirow{2}{*}{\textbf{Dataset}} & \multirow{2}{*}{\textbf{Artifact}} & \textbf{SGD} & \textbf{MAML} & \textbf{\begin{tabular}[c]{@{}c@{}}CMAML\\ (proposed)\end{tabular}} \\ \cline{3-5} 
 & & \textbf{PSNR/ SSIM} & \textbf{PSNR/ SSIM} & \textbf{PSNR/ SSIM} \\ \hline
\multirow{7}{*}{\begin{tabular}[c]{@{}c@{}}M\&M\\ \cite{campello2021multi}\end{tabular}} & Motion & 27.47/ 0.9043 & 27.64/ 0.9068 & \textbf{27.75/ 0.9079} \\ \cline{2-5} 
 & Gamma & 26.25/ 0.9610 & 26.67/ 0.9659 & \textbf{27.26/ 0.9681} \\ \cline{2-5} 
 & Ghosting & 31.10/ 0.9457 & 31.17/ 0.9470 & \textbf{31.18/ 0.9471} \\ \cline{2-5} 
 & Spiking & 27.93/ 0.8938 & 28.30/ 0.9004 & \textbf{28.40/ 0.9010} \\ \cline{2-5} 
 & US & 27.33/ 0.8853 & 27.32/ 0.8859 & \textbf{27.34/ 0.8862} \\ \cline{2-5} 
 & Noise & \textbf{28.59}/ 0.8748 & 28.26/ 0.8663 & 28.54/ \textbf{0.8749} \\ \cline{2-5} 
 & No artifact & 36.73/ 0.9922 & 37.04/ 0.9928 & \textbf{37.68/ 0.9930} \\ \hline
\multirow{7}{*}{\begin{tabular}[c]{@{}c@{}}ACDC\\ \cite{bernard2018deep}\end{tabular}} & Motion & 29.30/ 0.9282 & 29.54/ 0.9305 & \textbf{29.64/ 0.9313} \\ \cline{2-5} 
 & Gamma & \textbf{26.57}/ 0.9480 & 27.06/ 0.9526 & 27.56/ \textbf{0.9552} \\ \cline{2-5} 
 & Ghosting & 32.42/ 0.9639 & \textbf{32.51}/ 0.9655 & 32.45/ \textbf{0.9656} \\ \cline{2-5} 
 & Spiking & 29.77/ 0.9171 & \textbf{30.12/ 0.9218} & \textbf{30.12}/ 0.9212 \\ \cline{2-5} 
 & US & 25.81/ 0.8376 & \textbf{25.83}/ 0.8392 & 25.81/ \textbf{0.8393} \\ \cline{2-5} 
 & Noise & \textbf{29.92}/ 0.8765 & 29.66/ 0.8693 & 29.86/ \textbf{0.8771} \\ \cline{2-5} 
 & No artifact & 35.38/ 0.9871 & 36.23/ \textbf{0.9897} & \textbf{36.48/ 0.9897} \\ \hline
\end{tabular}
\end{table}
\begin{table}[t!]
\scriptsize
\centering
\caption{Quantitative results of composite artifacts on M\&M dataset. Bold figures indicate the best scores.}
\label{Compositional tabulation results}
\begin{tabular}{|c|c|c|c|}
\hline
\multirow{2}{*}{Composition of artifacts} & SGD & MAML & CMAML \\ \cline{2-4} 
 & PSNR/ SSIM & PSNR/ SSIM & PSNR/ SSIM \\ \hline
Noise+Spiking & 27.48/ 0.8897 & 27.52/ 0.8883 & \textbf{27.60/ 0.8904} \\ \hline
Under-sampling+Spiking & 26.70/ 0.8600 & 26.71/ 0.8612 & \textbf{26.73/ 0.8613} \\ \hline
Spatial-resolution+Noise & \textbf{29.13/ 0.8772} & 28.77/ 0.8692 & 28.90/ 0.8750 \\ \hline
Ghosting+Spiking & 27.21/ 0.9055 & \textbf{27.39}/ 0.9106 & 27.20/ \textbf{0.9110} \\ \hline
Under-sampling+Noise & 25.18/ 0.7988 & 25.08/ 0.7944 & \textbf{26.13/ 0.8004} \\ \hline
\end{tabular}
\end{table}

\section{EXPERIMENTAL DATASET DETAILS}
\label{sec:Data details}

\subsection{Training Data}
For training, we use Multi-Centre, Multi-Vendor \& Multi-Disease (M\&M) cardiac dataset \cite{campello2021multi}.
The train tasks (\( \mathcal{T} \)) include artifacts caused by motion, under-sampling and poor spatial resolution. We consider three motion artifact tasks with $s = 1, 2$ and $3$ in $\mathcal{A}_{M}$ operator, three under-sampling artifact tasks of $2x, 4x\ and\ 6x$ acceleration with Cartesian masks, and three scale factors of $2x, 3x\ and\ 5x$ as super-resolution tasks. In each task, we consider 2100 and 500 cardiac images for training and validation respectively. 
\subsection{Unseen Artifact Data}
\label{Unseen artifact data details}
We consider the following unseen artifact data for evaluation: (i) different cardiac dataset i.e., \textit{Automated Cardiac Diagnosis Challenge} (ACDC) \cite{bernard2018deep} dataset (ii) unseen amount of degradation (iii) unseen artifact type. The combination of three cases is also present in unseen artifact data.
Regarding the different degradations, we choose $s = 4$ and $6$ in $\mathcal{A}_{M}$ operator. Similarly, for undersampling artifact, we choose $3x$ and $5x$ acceleration in unseen artifact data. Moreover, artifacts not present in training data such as Spiking, Respiratory/Ghosting, Noise and Gamma artifacts are also present in unseen artifact data, simulated using \emph{TorchIO} \cite{perez2021torchio} library.
\subsection{Composite Artifact Data}
\label{Comp artifact data details}
Composite artifact data is characterized by simulating one artifact followed by a different artifact in the same cardiac scan. We consider five composite artifacts on M\&M data: 1) Noise and Spiking, 2) Spatial-resolution and Noise, 3) Under-sampling and Spiking, 4) Ghosting and Spiking, and 5) Under-sampling and Noise. Undersampling artifact consists of $3x$ acceleration and a $2x$ factor for spatial resolution.
\subsection{Preprocessing}
M\&M and ACDC dataset images consist of three spatial dimensions and one time dimension, along with segementation maps and end-systolic (ES) \cite{fukuta2008cardiac} time instance of the cardiac cycle. In the time dimension, except the first and last time points, the rest of the slices are considered for training. The segmentation map of the ES slice consists of the clinically relevant portion of the cardiac scan. A 128x128 region is cropped around the region of interest defined by the segmentation map. Images are normalized for the intensity homogeneity across CMR scans. 

\section{IMPLEMENTATION DETAILS}
\label{sec:implementatin details}
The DL model is a five layer CNN \cite{wang}. Learning rates $(\alpha, \beta)$ of 0.001 with 200 $max\_epochs$ are chosen. Outer level and inner level use Adam and SGD optimizers respectively. Batch size for support and query data $(N_{spt}, N_{qry})$ is 5. Task mini-batch $(|d|)$ is chosen to be 3. Pacing functions's cumulative inclusion of artifact data across epochs is shown in Fig. \ref{fig:Pacing function} along with easy, medium and hard artifacts. Pacing-function also modulates the number of adaptation steps $(U)$ from $1, 2\ \&\ 3$ at the cue of adding graded artifacts during training. The loss function is $L_{1}$ norm between the model's prediction and ground truth. All models are trained on $24$GB memory \emph{Nvidia RTX-3090}.

\section{RESULTS AND DISCUSSION}
\label{sec:results}
\subsection{Evaluation on unseen artifacts}
MRI modality is diverse due to its varied acquisition settings and patient conditions. We evaluate our proposed method under deviated scenario of different train and test artifacts and degradations. Details on the unseen artifact data that is used to qualitatively and quantitatively demonstrate the CMAML generalization is given in Section \ref{Unseen artifact data details}. Table \ref{Tabulation results} provides quantitative results in terms of SSIM and PSNR (dB) \cite{sara2019image} metrics. From Table \ref{Tabulation results}, for M\&M dataset, CMAML is consistently better than other methods in terms of SSIM metric that is considered to be close to radiologist scores \cite{Muckley2021ResultsOT}. Similarly, in terms of PSNR, CMAML performs better artifact suppression for 83\% of unseen degradations on M\&M dataset. Likewise, on ACDC dataset, meta-learning methods outperform SGD for 83\% and 67\% of artifacts in terms of SSIM and PSNR respectively. Maximum improvement is present in spiking and gamma artifacts in M\&M and ACDC dataset respectively. Qualitative results with residue images in Fig. \ref{fig:Noise artifact visualization} indicate that the proposed CMAML is robust with respect to cardiac image perturbations than other methods. Artifact restoration of motion affected image shown in Fig. \ref{fig:Motion artifact visualization} indicate better recovery of details in left ventricle enclosed by a yellow bounding box in residue. From the boxplot of various artifacts in Fig. \ref{SSIM boxplot}, we believe that the regularization offered by curriculum in MAML framework better aids the restoration of CMR images, especially in spiking and additive noise corruption. For an input image with no artifact, CMAML shows better identity mapping over other methods as shown in Table. \ref{Tabulation results}.
\subsection{Evaluation on composite artifacts}
The scenario of composition of artifacts where the scan contains multiple degradations is prominent in MRI acquisition \cite{zaitsev2015motion}. For instance, the ghosting artifact from the respiratory cardiac movement is compounded by the under-sampling artifact arising from acquisition settings. We evaluate on the test scenario that contains two different artifacts in the same scan in Table \ref{Compositional tabulation results}. Details on the composite artifact data is given in Section \ref{Comp artifact data details}. From Table \ref{Compositional tabulation results}, our CMAML method is quantitatively performing better than joint training and MAML in terms of SSIM scores for 80\% of composite artifacts. Similarly, in terms of PSNR metric, meta learning methods outperform SGD for 80\% of composite artifacts. For the composite artifact of \emph{ghosting+spiking}, a maximum improvement margin of around 0.006 is noticed for CMAML over SGD. Qualitative performance of various methods is shown in Fig. \ref{fig:Composite artifact visualization}, including the residue, for a combination of undersampling followed by additive Gaussian noise. The clinically prominent portion of the left ventricle of cardiac is highlighted by a yellow bounding box in the residue images of Fig. \ref{fig:Composite artifact visualization} and CMAML predicts the artifact suppressed image closer to the normal cardiac scan. For the proposed method, the metrics are statistically significant with $p < 0.05$.

\section{CONCLUSIONS}
We propose CMAML, a training process based on meta-learning, to efficiently consolidate the information of multiple types of artifacts for MRI artifact image restoration. The training process is enabled to embed better representations by associating the artifact's ill-posedness to the complexity and thus effectively promoting the curriculum fashion of training into meta-learning framework. For a fixed neural network, we demonstrated the better generalizability of CMAML on various nuances of unseen artifacts and composite artifacts in cardiac MRI images qualitatively and quantitatively.

Our model has been trained with deployment perspectives to improve diagnostic quality under various unseen artifacts, and has the potential to eliminate models trained separately for each artifact. Moreover, restoring the artifact affected scan within the MRI workflow can reduce the burden on healthcare system by avoiding patient recall and rescan. Future work is to incorporate self-supervision into CMAML.

\addtolength{\textheight}{-12cm}   
\bibliographystyle{IEEEbib}
\bibliography{refs}

\begin{thebibliography}{10}

\bibitem{ismail2022cardiac}
T.F. Ismail et~al.,
\newblock ``Cardiac mr: from theory to practice,''
\newblock {\em Frontiers in cardiovascular medicine}, vol. 9, 2022.

\bibitem{cs_mri}
M.~Lustig et~al.,
\newblock ``Sparse {MRI}: The application of compressed sensing for rapid {MR}
  imaging,''
\newblock {\em Magn. Reson. Med.}, 2007.

\bibitem{Kstner2020CINENetDL}
T.~K{\"u}stner et~al.,
\newblock ``Cinenet: deep learning-based 3d cardiac cine mri reconstruction
  with multi-coil complex-valued 4d spatio-temporal convolutions,''
\newblock {\em Scientific Reports}, vol. 10, 2020.

\bibitem{lyu2021cine}
Q.~Lyu et~al.,
\newblock ``Cine cardiac mri motion artifact reduction using a recurrent neural
  network,''
\newblock {\em IEEE Transactions on Medical Imaging}, vol. 40, no. 8, pp.
  2170--2181, 2021.

\bibitem{xia2021super}
Y.~Xia et~al.,
\newblock ``Super-resolution of cardiac mr cine imaging using conditional gans
  and unsupervised transfer learning,''
\newblock {\em Medical Image Analysis}, vol. 71, pp. 102037, 2021.

\bibitem{masutani2020deep}
E.M. Masutani et~al.,
\newblock ``Deep learning single-frame and multiframe super-resolution for
  cardiac mri,''
\newblock {\em Radiology}, vol. 295, pp. 552, 2020.

\bibitem{univusmri}
X.~Liu et~al.,
\newblock ``Universal undersampled mri reconstruction,''
\newblock in {\em International Conference on Medical Image Computing and
  Computer-Assisted Intervention}. Springer, 2021, pp. 211--221.

\bibitem{ramanarayanan2020mac}
S.~Ramanarayanan et~al.,
\newblock ``Mac-reconnet: A multiple acquisition context based convolutional
  neural network for mr image reconstruction using dynamic weight prediction,''
\newblock in {\em Medical Imaging with Deep Learning}. PMLR, 2020, pp.
  696--708.

\bibitem{finn2017model}
C.~Finn et~al.,
\newblock ``Model-agnostic meta-learning for fast adaptation of deep
  networks,''
\newblock in {\em International conference on machine learning}. PMLR, 2017,
  pp. 1126--1135.

\bibitem{hacohen2019power}
G.~Hacohen et~al.,
\newblock ``On the power of curriculum learning in training deep networks,''
\newblock in {\em International Conference on Machine Learning}. PMLR, 2019,
  pp. 2535--2544.

\bibitem{yang2017dagan}
G.~Yang et~al.,
\newblock ``Dagan: deep de-aliasing generative adversarial networks for fast
  compressed sensing mri reconstruction,''
\newblock {\em IEEE transactions on medical imaging}, vol. 37, no. 6, pp.
  1310--1321, 2017.

\bibitem{hospedales2021meta}
T.~Hospedales et~al.,
\newblock ``Meta-learning in neural networks: A survey,''
\newblock {\em IEEE transactions on pattern analysis and machine intelligence},
  2021.

\bibitem{campello2021multi}
V.~M. Campello et~al.,
\newblock ``Multi-centre, multi-vendor and multi-disease cardiac segmentation:
  the m\&ms challenge,''
\newblock {\em IEEE Transactions on Medical Imaging}, vol. 40, no. 12, pp.
  3543--3554, 2021.

\bibitem{bernard2018deep}
O.~Bernard et~al.,
\newblock ``Deep learning techniques for automatic mri cardiac multi-structures
  segmentation and diagnosis: is the problem solved?,''
\newblock {\em IEEE transactions on medical imaging}, vol. 37, pp. 2514--2525,
  2018.

\bibitem{perez2021torchio}
F.~P{\'e}rez-Garc{\'\i}a et~al.,
\newblock ``Torchio: a python library for efficient loading, preprocessing,
  augmentation and patch-based sampling of medical images in deep learning,''
\newblock {\em Computer Methods and Programs in Biomedicine}, vol. 208, pp.
  106236, 2021.

\bibitem{fukuta2008cardiac}
F.~Hidekatsu et~al.,
\newblock ``The cardiac cycle and the physiologic basis of left ventricular
  contraction, ejection, relaxation, and filling,''
\newblock {\em Heart failure clinics}, vol. 4, no. 1, pp. 1--11, 2008.

\bibitem{wang}
S.~{Wang} et~al.,
\newblock ``Accelerating magnetic resonance imaging via deep learning,''
\newblock in {\em International Symposium on Biomedical Imaging}, 2016.

\bibitem{sara2019image}
S.~Umme et~al.,
\newblock ``Image quality assessment through fsim, ssim, mse and psnr—a
  comparative study,''
\newblock {\em Journal of Computer and Communications}, vol. 7, no. 3, pp.
  8--18, 2019.

\bibitem{Muckley2021ResultsOT}
M.~Muckley et~al.,
\newblock ``Results of the 2020 fastmri challenge for machine learning mr image
  reconstruction,''
\newblock {\em IEEE transactions on medical imaging}, vol. 40, pp. 2306 --
  2317, 2021.

\bibitem{zaitsev2015motion}
Z.~Maxim et~al.,
\newblock ``Motion artifacts in mri: A complex problem with many partial
  solutions,''
\newblock {\em Journal of Magnetic Resonance Imaging}, vol. 42, no. 4, pp.
  887--901, 2015.

\end{thebibliography}

\end{document}